\documentclass[a4paper]{article}

\usepackage{blindtext}
\usepackage[latin1]{inputenc}
\usepackage[T1]{fontenc}
\usepackage[english]{babel}
\usepackage{amsmath} 
\usepackage{amssymb}
\usepackage{array}
\usepackage{color}
\usepackage{enumerate}
\usepackage{fancybox}
\usepackage{float}
\usepackage[hang]{footmisc}
\usepackage{graphicx}
\usepackage{lmodern}
\usepackage{multirow}
\usepackage{pifont}
\usepackage{rotating}      
\usepackage[small, sl, SL,sf, SF, bf, hang, raggedright]{subfigure}
\usepackage{tcolorbox}
\usepackage{textpos}
\usepackage{tikz}
\usepackage{ulem} 
\usepackage{upgreek}

\usetikzlibrary{arrows,shapes,positioning}
\usetikzlibrary{decorations.markings}
\usetikzlibrary{decorations.pathmorphing}
\usetikzlibrary{patterns}
\usetikzlibrary{plotmarks}
\usetikzlibrary{shadows}

\tcbuselibrary{skins}

\tikzstyle arrowstyle=[scale=1.5]
\tikzstyle directed=[postaction={decorate,decoration={markings,
    mark=at position .55 with {\arrow[arrowstyle]{stealth}}}}]
\tikzstyle reverse directed=[postaction={decorate,decoration={markings,
    mark=at position .55 with {\arrowreversed[arrowstyle]{stealth};}}}]
\tikzset{
    boson/.style={decorate, decoration={snake}, draw=black},
    fermion/.style={draw=black, postaction={decorate},
        decoration={markings,mark=at position .6 with {\arrow[arrowstyle]{stealth}}}},
    fbar/.style={draw=black, postaction={decorate},
        decoration={markings,mark=at position .6 with {\arrowreversed[arrowstyle]{stealth}}}},
    gluon/.style={decorate, draw=black,
        decoration={coil,amplitude=4pt, segment length=5pt}} 
}

\let\beginoldtabular=\tabular
\let\endoldtabular\endtabular
\renewenvironment{tabular}
{	
	
	\beginoldtabular
}
{
	\endoldtabular
}

\usepackage{authblk}

\usepackage[hmarginratio=1:1,top=32mm,left=20mm,right=20mm,columnsep=20pt]{geometry} 
\usepackage{multicol} 
\usepackage[hang, small,labelfont=bf,up,textfont=it,up]{caption} 
\usepackage{booktabs} 
\usepackage{float} 
\usepackage{hyperref} 

\usepackage{abstract} 

\usepackage{titlesec} 
\titleformat{\section}[block]{\large\scshape\centering}{\thesection.}{1em}{} 
\titleformat{\subsection}[block]{\large}{\thesubsection.}{1em}{} 

\usepackage{fancyhdr} 
\makeatletter
\def\@maketitle{%
  \begin{center}%
  \let \footnote \thanks
  	\vspace{-1cm}
    {\huge \textbf{\@title} \par}%
    \vspace{0.8cm}%
    {\large \@author \par  }%
    \vspace{0.5cm}%
   { \@date}%
   \vspace{0.25cm}%
  \end{center}%
 }
\makeatother

\title{Polarimetry at the ILC}    

\author[,1,2,3]{Robert Karl\thanks{email: robert.karl@desy.de}}
\author[,1]{Jenny List\thanks{email: jenny.list@desy.de}}
\affil[1]{Deutsches Elektronen-Synchrotron (DESY)}
\affil[2]{University of Hamburg, Germany}
\date{06.12.16}

\begin{document}

\maketitle 
\thispagestyle{empty} 

\begin{abstract}
At the ILC, the luminosity-weighted average polarization at the IP needs to be determined at the permille-level. In order to reach this goal, the combined information from the polarimeter and the collision data is required. In this study, a unified approach will be presented, which for the first time combines the cross section measurements with the expected constraints from the polarimeters. Hereby, the statistical and systematical uncertainties are taken into account, including their correlations.\\
This study shows that a fast spin flip frequency is required because it easily reduces the systematic uncertainty, while a non-perfect helicity reversal can be compensated for within the unified approach. The final goal is to provide a realistic estimation of the luminosity-weighted average polarization at the IP to be used in the physic analyses.
\end{abstract}
\vfill
\begin{center}
\large{\footnotemark[3] Talk presented at the International Workshop on Future Linear Colliders\\[0.5em] (LCWS2016),\\[0.5em] Morioka, Japan, 5-9 December 2016. C16-12-05.4.}
\end{center}
\vfill

\newpage

\section{Introduction}

The usage of polarized beams provides great advantages for the International Linear Collider (ILC). It allows deep insights into the chiral structure of the weak-interaction for known and unknown particle as well as a sensitivity to additional observables (e.g. left-right-asymmetry). Furthermore, with polarized beams it is possible to suppress background processes and simultaneously increase signal processes, providing an enhancement of the signal to noise ratio.\\
Thus, the ILC\cite{ILCProjectVol1} beams will be polarized to a degree of $\left|80\%\right|$ for the electron beam and $\left|30\%\text{ - }60\%\right|$ for the positron beam. The sign of the polarization is individually adjustable for each beam, providing a choice of different spin configurations. However, since all event rates depend linearly on the polarization, it is important to provide a determination of the actual beam polarization at the permille-level in order to fully exploit the physics potential of the ILC\cite{PolarizationRequirement}. This requirement can only be fulfilled by combining the fast time-resolved measurements of the laser-Compton polarimeters with an absolute scale calibration of the luminosity-weighted average polarization at the interaction point (IP) calculated from collision data.
\begin{figure}[htbp]
\includegraphics[width = \textwidth]{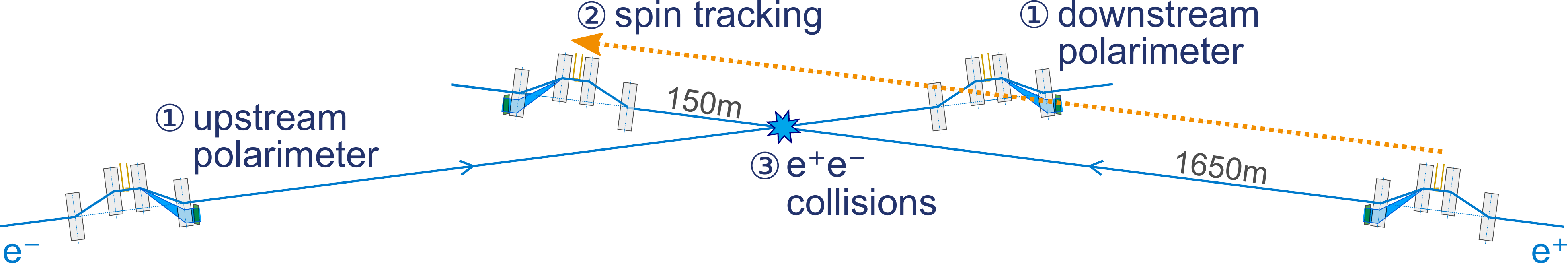}
\caption{
Sketch of the ILC polarimetry concept: \ding{172} The beam polarization will be measured online by 2 laser-Compton polarimeters~\cite{polarimeter} per beam. For each beam, the upstream polarimeter is located 1.65\,km before the IP, while the downstream polarimeter is located 150\,m after the IP. \ding{173} In order to determine the polarization at the IP, the result of the polarimeter measurement is extrapolated to the IP via spin tracking~\cite{spintracking}. \ding{174} For the absolute scale calibration, the luminosity-weighed averaged polarization is additionally calculated from collision data.
}
\label{fig:PolarimeterConcept}
\end{figure}

As shown in Fig.~\ref{fig:PolarimeterConcept}, the time-resolved polarization is measured by two polarimeters per beam from the differential Compton cross section of a particle bunch with a circular polarized lasers~\cite{polarimeter}. However, those polarimeters are 1.65\,km before and 150\,m after the IP. Thus, the measurement has to be extrapolated from the polarimeters to the IP via spin tracking~\cite{spintracking}, considering all uncertainties including beam collision effects. The absolute scale calibration is determined by the luminosity-weighed averaged polarization calculated from the cross section measurement from different, well known standard model processes.\\
In the past, such a polarization determination from collision data has been studied for different standard model processes individually. In previous studies, the polarization was determined using the information from W-pair production~\cite{wpairstudy} and using the information from single $W$, $\gamma$, $Z$ events~\cite{singlebosonstudy}. However, the current goal is to find a general strategy for the polarization determination which yields the best precision per measurement time. Thus, the following criteria have to be considered:\\
\begin{itemize}
	\item \textbf{Combining all relevant processes}\\
	By combining all suitable processes, the statistical precision can be increased and different systematic uncertainties give a better control. Suitable processes exhibit a large left-right-asymmetry and a high cross section. For a precise determination all uncertainties and their correlations have to be taken into account. This also provides a more robust polarization determination against possible BSM effects of single processes.\\
	\item \textbf{Compensating for a non-perfect helicity reversal}\\
	In realistic running condition, the absolute polarization will also slightly change when reversing the sign. This has to be considered for the polarization determination in order to increase its precision.\\\newpage
	\item \textbf{Including constraints from the polarimeter measurement}\\
	Since the polarimeter already yields a polarization measurement, it can be used as a constraint for the calculation of the cross section measurement to make it more robust against large statistical fluctuations, especially for low luminosities. For a precise determination, all sources of uncertainties, e.g. from spin tracking, have to be taken into account.\\
\end{itemize}

\section{Polarization Measurement Using Collision Data}

In order to determine the polarization from collision data, the theoretically predicted cross section of each process is compared to the corresponding measurement within the uncertainty by variating the beam polarization equally for all four spin configurations. For this purpose, a $\chi^{2}$-minimization method is used.

\subsection{Combination of Cross Section Measurements}

In order to combine the cross section measurement, the $\chi^{2}$ values of the individual processes are summed up and share a common polarization parameter set. To consider all systematic uncertainties and their correlation, a full covariance matrix $\Xi$ was used, shown in eq.~\ref{eq:chi2formula}. This $\chi^{2}$-function will be refered to \textit{the unified approach}.

\begin{align}
\chi^{2} &:=
\sum_{\text{process}}{
\left(\vec{\sigma}_{\text{data}} - \vec{\sigma}_{\text{theory}}\right)^{T}\Xi^{-1}
\left(\vec{\sigma}_{\text{data}} - \vec{\sigma}_{\text{theory}}\right)
};&\qquad
\vec{\sigma} &:= \begin{pmatrix}
\sigma_{-+} & \sigma_{+-} & \sigma_{--} & \sigma_{++}\\
\end{pmatrix}^{T}
\label{eq:chi2formula}
\end{align}

Here, $\Xi^{-1}$ refers to the inverse of the covariance matrix. The covariance matrix is derived from the error propagation of the cross section measurement, as shown in eq.~\ref{eq:covariancematrix}.

\begin{align}
\left(\vec{\sigma}_{\text{data}}\right)_{i} &= \frac{D_{i} - \mathfrak{B}_{i}}{\varepsilon_{i}\cdot\mathcal{L}_{i}}
&\Rightarrow\quad\textbf{e.g.: }\left(\Xi_{\varepsilon}\right)_{ij} &= 
	\text{corr}\left(\vec{\sigma}_{i}^{\varepsilon},\ \vec{\sigma}_{j}^{\varepsilon}\right)
	\frac{\partial\vec{\sigma}_{i}}{\partial\varepsilon_{i}}\frac{\partial\vec{\sigma}_{j}}{\partial\varepsilon_{j}}
	\Delta\varepsilon_{i}\Delta\varepsilon_{j}
&\Rightarrow\quad\Xi &:= \Xi_{D} + \Xi_{\mathfrak{B}} + \Xi_{\varepsilon} + \Xi_{\mathcal{L}};
\label{eq:covariancematrix}
\end{align}

Here, $D_{i}$ is the number of signal events, $\mathfrak{B}_{i}$ is the background expectation value, $\varepsilon_{i}$ is the selection efficiency of the detector and $\mathcal{L}_{i}$ is the integrated luminosity of the data set. The index $i$ corresponds to the different helicity configurations $\left(-+,\ +-,\ --,\ ++\right)$. The final covariance matrix is the sum of the individual covariance matrices of the four quantities ($D_{i}$, $\mathfrak{B}_{i}$, $\varepsilon_{i}$, $\mathcal{L}_{i}$). The correlation factors (e.g. $\text{corr}\left(\vec{\sigma}_{i}^{\varepsilon},\ \vec{\sigma}_{j}^{\varepsilon}\right)$) can vary for each quantity but is equal for each process. Furthermore, only correlations between the different spin configurations are considered individually for each quantity. Correlations between the different quantities are neglected. The correlation factors are fixed and have to determined externally. Since the signal $D_{i}$ is always uncorrelated, $\Xi_{D}$ is here always a diagonal matrix.\\
Compensation for a non-perfect helicity reversal is achieved by treating the polarization value for both helicities and beams as independent. This results in 4 free parameters defined with their nominal values in eq.~\ref{eq:freeparameters}.

\begin{align}
\underbrace{P_{e^{-}}^{-} = -80\%,}_{\text{"left"-handed }e^{-}\text{-beam}}\qquad
\underbrace{P_{e^{-}}^{+} =  80\%,}_{\text{"right"-handed }e^{-}\text{-beam}}\qquad
\underbrace{P_{e^{+}}^{-} = -30\%,}_{\text{"left"-handed }e^{+}\text{-beam}}\qquad
\underbrace{P_{e^{+}}^{+} =  30\%,}_{\text{"right"-handed }e^{+}\text{-beam}}
\label{eq:freeparameters}
\end{align}

An alternative parametrization, which was used in previous studies, is to use an absolute average polarization between both helicities and the deviations, as defined in eq.~\ref{eq:alternativeparameters}. Both parametrizations are equivalent but for this study the parametrization defined in eq.~\ref{eq:freeparameters} was used because it has an advantage by using the polarimeter constraint, as described later.

\begin{align}
	P_{e^{\pm}}^{-} & = -\left|P_{e^{\pm}}\right| + \tfrac{1}{2}\delta_{e^{\pm}} &  
	P_{e^{\pm}}^{+} & = \quad\left|P_{e^{\pm}}\right| + \tfrac{1}{2}\delta_{e^{\pm}}
	\label{eq:alternativeparameters}
\end{align}

The processes used for the polarization calculation are listed in tab.~\ref{tab:Consideredprocesses}. They were selected due to their relatively large left-right-asymmetry and unpolarized cross section in order to gain the most sensitive processes to the polarization with the highest event rate. Thereby, they are the same processes as used for physics analyses (DBD) with respect to their classification, labeling and cross section values. The chiral cross section of those processes were calculated on tree-level but including ISR to consider the reduced center-of-mass energy. Furthermore, any combination of processes can be used to study the effect of different combinations and the sensitivity of the different processes on the polarization precision. Thereby, further processes can easily be added. 

\begin{table}[htbp]
\centering
\begin{tabular}{|c|c|c|c|c|c|}\hline
\textbf{Process} & single$W^{\pm}$ & $WW$ & $ZZ$ & $ZZWW$Mix &$Z$ \\\hline
\textbf{Channel} & $e\nu q\bar{q}$, $e\nu l\nu$ & $q\bar{q}q\bar{q}$, $q\bar{q}l\nu$, $l\nu l\nu$ & $q\bar{q}q\bar{q}$, $q\bar{q}ll$,$llll$ & $q\bar{q}q\bar{q}$, $l\nu l\nu$ & $q\bar{q}$, $ll$ \\\hline
\end{tabular}
\caption{Currently considered processes which are suitable for minimization due to their left-right-asymmetry and unpolarized cross section.}
\label{tab:Consideredprocesses}
\end{table}

In order to test the theoretical limit on the polarization precision, a perfect $4\pi$ detector is assumed with a zero background estimation value and no systematic uncertainties. The statistical precision limit for such a scenario is shown in tab.~\ref{tab:statisticalprecision}\,(\textit{left}) for the H-20\cite{ILCrunningscenario} scenario using all implemented processes listed in tab.~\ref{tab:Consideredprocesses}. For the data sets with larger integrated luminosity, the permille-level is clearly achievable but for lower integrated luminosity data sets (e.g. top-threshold scan at 350\,GeV), this does not apply. However, in order to still achieve the permille-level precision in particular for lower integrated luminosities, the polarimeter constraint will become important, as described in sec.~\ref{sec:polarimeterconstraint}.

\begin{table}[htbp]
\begin{tabular}{|c|c|c|c||c|c|}\hline
\multicolumn{6}{|c|}{\textbf{The Unified Approach}}\\\hline
$E$[GeV] & $\textbf{500}$ & $\textbf{350}$ & $\textbf{250}$ & $\textbf{500}$ & $\textbf{250}$\\\hline
$\mathcal{L}$[fb$^{-1}$] & $\textbf{500}$ & $\textbf{200}$ & $\textbf{500}$ & $\textbf{3500}$ & $\textbf{1500}$ \\\hline
$\Delta P_{e^{-}}^{-}/P$ & \textcolor{blue}{$0.2$} & \textcolor{red}{$0.3$} & $0.1$ & $0.08$ & $0.09$\\\hline
$\Delta P_{e^{-}}^{+}/P$ & $0.05$ & $0.06$ & $0.03$ & $0.02$ & $0.02$\\\hline
$\Delta P_{e^{+}}^{-}/P$ & $0.1$ & $0.1$ & $0.06$ & $0.04$ & $0.04$\\\hline
$\Delta P_{e^{+}}^{+}/P$ & \textcolor{blue}{$0.2$} & \textcolor{red}{$0.3$} & $0.1$ & $0.08$ & $0.08$\\\hline
\end{tabular}\hfill
\begin{tabular}{|c|c|}\hline
\multicolumn{2}{|c|}{\textbf{$W$-Pair}~\cite{wpairstudy}}\\\hline
$E$[GeV] & $\textbf{500}$\\\hline
$\mathcal{L}$[fb$^{-1}$] & $\textbf{500}$\\\hline
\multirow{2}{*}{$\Delta P_{e^{-}}/P$}	&	 \multirow{2}{*}{$0.08$	}\\
&\\\hline
\multirow{2}{*}{$\Delta P_{e^{+}}/P$}	&	 \multirow{2}{*}{$0.34$	}\\ 
&\\\hline
\end{tabular}\hfill
\begin{tabular}{|c|c|}\hline
\multicolumn{2}{|c|}{\textbf{Single Boson}~\cite{singlebosonstudy}}\\\hline
$E$[GeV] & $\textbf{500}$\\\hline
$\mathcal{L}$[fb$^{-1}$] & $\textbf{2000}$\\\hline
$\Delta P_{e^{-}}/P$		& $0.085$	\\\hline
$\Delta \delta_{e^{-}}/P$	& $0.12$	\\ \hline
$\Delta P_{e^{+}}/P$ 		& $0.22$	\\\hline
$\Delta \delta_{e^{+}}/P$	& $0.32$	\\\hline
\end{tabular}\hfill
\caption{The theoretical statistical precision $\Delta P / P\left[\%\right]$ limit for the unified approach using the H-20\cite{ILCrunningscenario} scenario in comparison to previous studies.}
\label{tab:statisticalprecision}
\end{table}

\begin{figure}[htbp]
\centering
\includegraphics[width=0.5\textwidth]{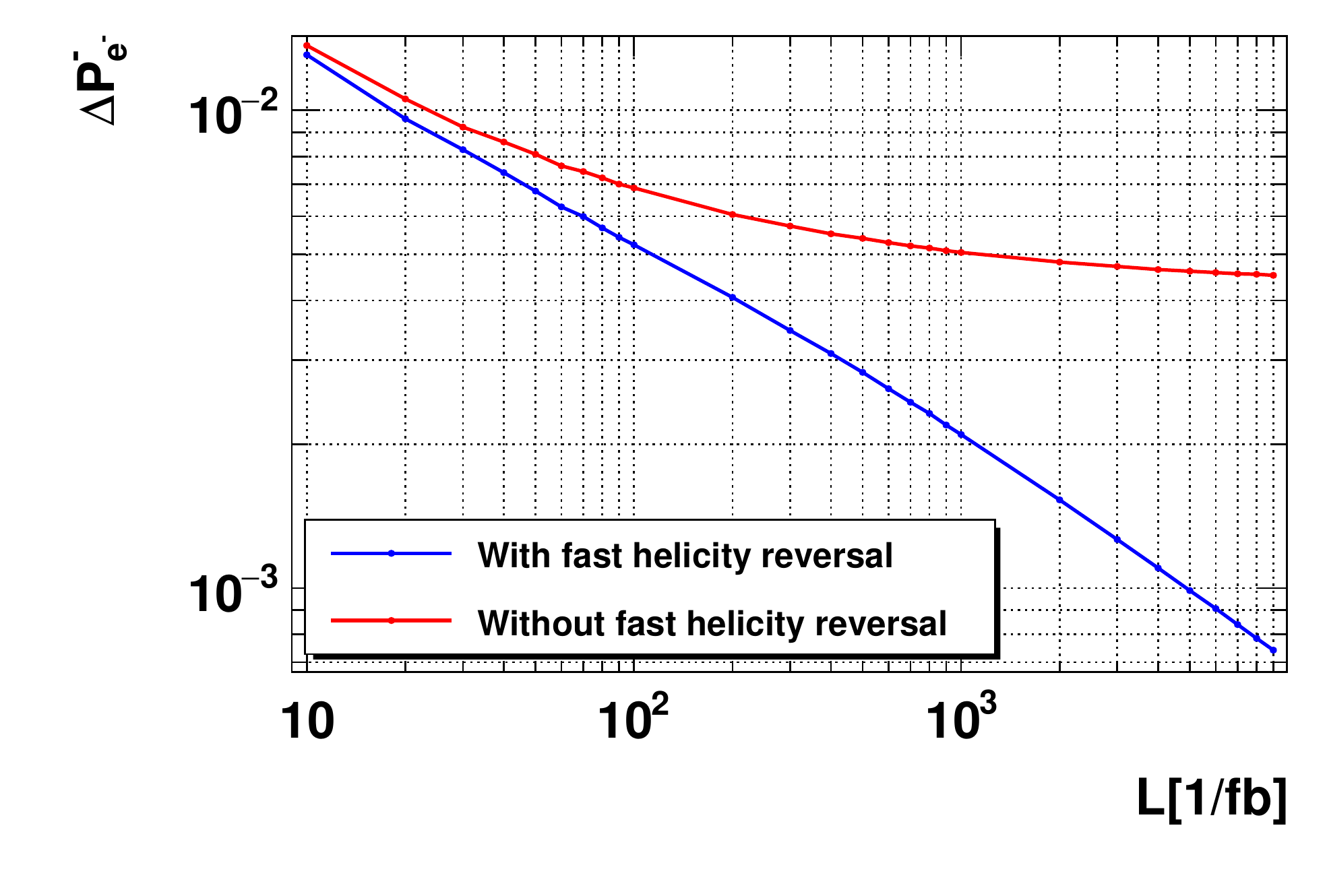}
\caption{The effects of the fast helicity reversal on the polarization precision because of correlated uncertainties.}
\label{fig:correlateduncertainty}
\end{figure}

In order to get a more realistic point of view on the achievable statistical precision, the results are shown in comparison to two previous studies. Both of them also use only statistical uncertainties and a $\chi^{2}$ minimization. The first is the polarization calculation from $W$-pairs, shown in the middle of tab.~\ref{tab:statisticalprecision}. It additionally uses the information from the production angle and includes fiducial cuts and a complete background calculation. However, for this study, only a 2 parameter fit was performed because a constant absolute polarization was assumed for the helicity reversal. The second study used a combined cross section measurement from single $W^{+}$, $W^{-}$, $Z$, $\gamma$ processes. No angular information was used and no background was considered for this study. But it included fiducial cuts on the total cross section and considered a deviation on the absolute polarization as free parameter. However, $\delta_{e^{\pm}}$ was constraint to $10^{-3}$ in this study. In comparison to the ideal case of the unified approach, the precision is in the same order of magnitude. It shows that even a more realistic scenario can get close on the statistic precision limit.\\
In order to reach the goal of 0.1\% on the total uncertainty, the systematical uncertainties also need to be controlled at the permille-level. The uncertainties on the integrated luminosity $\Delta\mathcal{L}$ and the selection efficiency $\Delta\varepsilon$ are influenced by the machine performance and detector calibration and alignment, respectively. Thus, $\Delta\mathcal{L}$ and $\Delta\varepsilon$ are time dependent. If the switch between the different spin configuration (helicity reversal) is faster than a change in the calibration, alignment, etc. of the detector and accelerator, $\Delta\mathcal{L}$ and $\Delta\varepsilon$ become correlated, as introduced in eq.~\ref{eq:covariancematrix}. This correlation leads to cancellation of the systematic uncertainties reducing the impact on the polarization precision. Otherwise, the polarization precision would saturate and it would be limited by the systematic uncertainties, as shown in Fig.~\ref{fig:correlateduncertainty}.\\
Therefore, a fast helicity reversal (e.g. train-by-train) is required to reduce the influence of the time dependent systematic uncertainties on the polarization precision. Note that this still applies even if the helicity reversal is not perfect because a non-perfect helicity reversal has close to no influence on the precision due to compensation of the unified approach.\\  

\subsection{Frequency and Accuracy of the Fast Helisity Reversal}

The advantage of a fast helicity reversal was demonstrated in the last section. For the realization, it is important that the switch between the four different spin configuration is performed during normal operation with no additional breaks. As seen in Fig.~\ref{fig:bunchstrcture} for the ILC bunch structure, there are in principle two possible options:

\begin{figure}[htbp]
\centering
\includegraphics[width=0.5\textwidth]{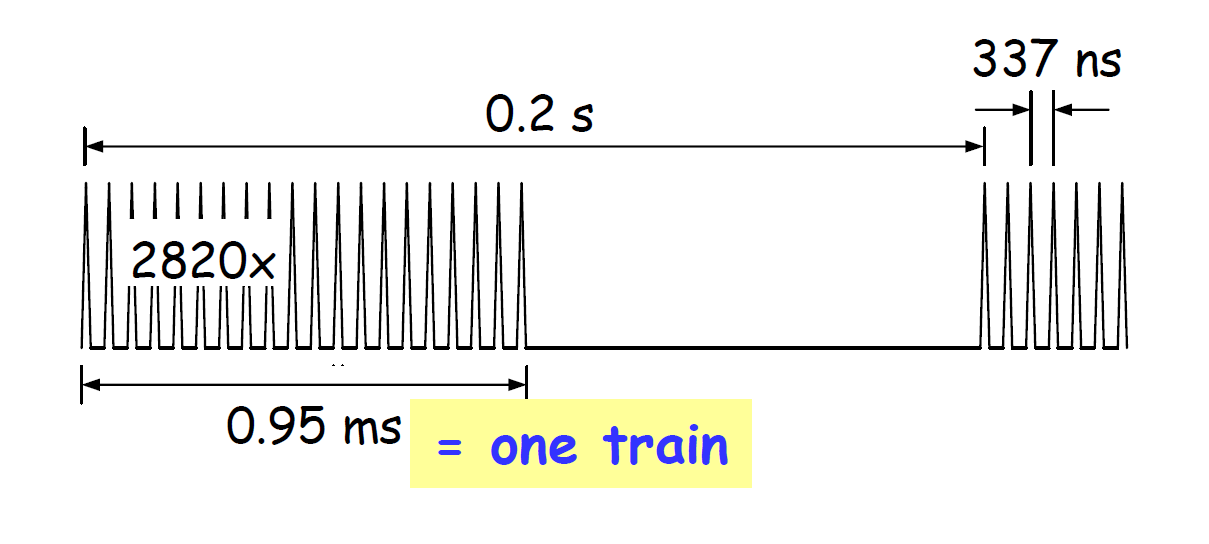}
\caption{The ILC structure of a bunch train}
\label{fig:bunchstrcture}
\end{figure}

\begin{itemize}
	\item\textbf{Bunch-by-bunch:}\\
	The helicity reversal is faster than the duration the 337\,ns time gap between two bunches.
	\item\textbf{Train-by-train:}\\
	The helicity reversal is performed within the break between two trains. In the ILC design, an approximately 199\,ms break is foreseen between the trains. This time slot, which is inter alia used for readout of the detector, can also be used to switch between the different spin configurations.  
\end{itemize}

\begin{figure}[htb]
\centering
\includegraphics[width=0.8\textwidth]{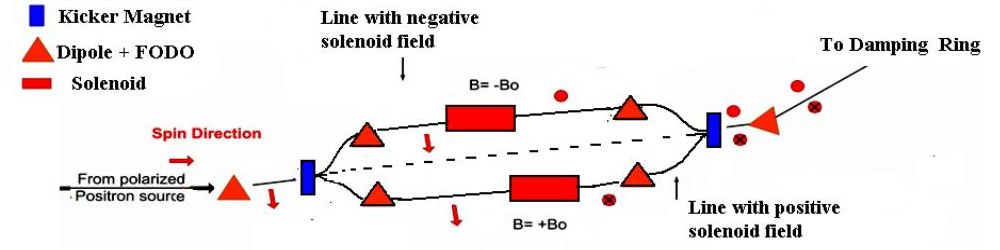}
\caption{The schematic layout of positron transport to Damping Ring with a two parallel lines spin rotator section.\cite{positronflip}}
\label{fig:positronflip}
\end{figure}

It is assumed that the switch train-by-train yields a sufficient correlations to suppress systematic uncertainty but a precise determination of the correlation factor is still an open topic. For a realistic scenario, a non-perfect helicity reversal has to be considered. This is particularly important for the positron beam. As shown in Fig.~\ref{fig:positronflip}, the helicity reversal for the positron beam is accomplished by switching between two beam lines, which also enables the possibility of a fast helicity reversal. The polarized electron beam is generated by shooting a circular polarized laser onto a photo cathode. The sign of the electron beam polarization corresponds to the sign of the laser polarization. Thus, the helicity reversal for the electron beam is achieved by a switch of the laser polarization, which can easily performed train-by-train. Here, the deviation from a perfect helicity reversal are assumed to be very small. In principle, a non-perfect helicity reversal is no issue as long as it can be accurately measured. 

To show that the new unified approach can handle variations in the absolute polarization, toy measurements with 5 different polarization discrepancies for both beams were performed. However, the nominal absolute polarization values $\left|P_{e^{-}}\right| = 80\%$, $\left|P_{e^{+}}\right| = 30\%$ were used as initial parameters for each toy measurements. Only statistical uncertainties were assumed.

\begin{figure}[htb]
\centering
\includegraphics[width=0.49\textwidth]{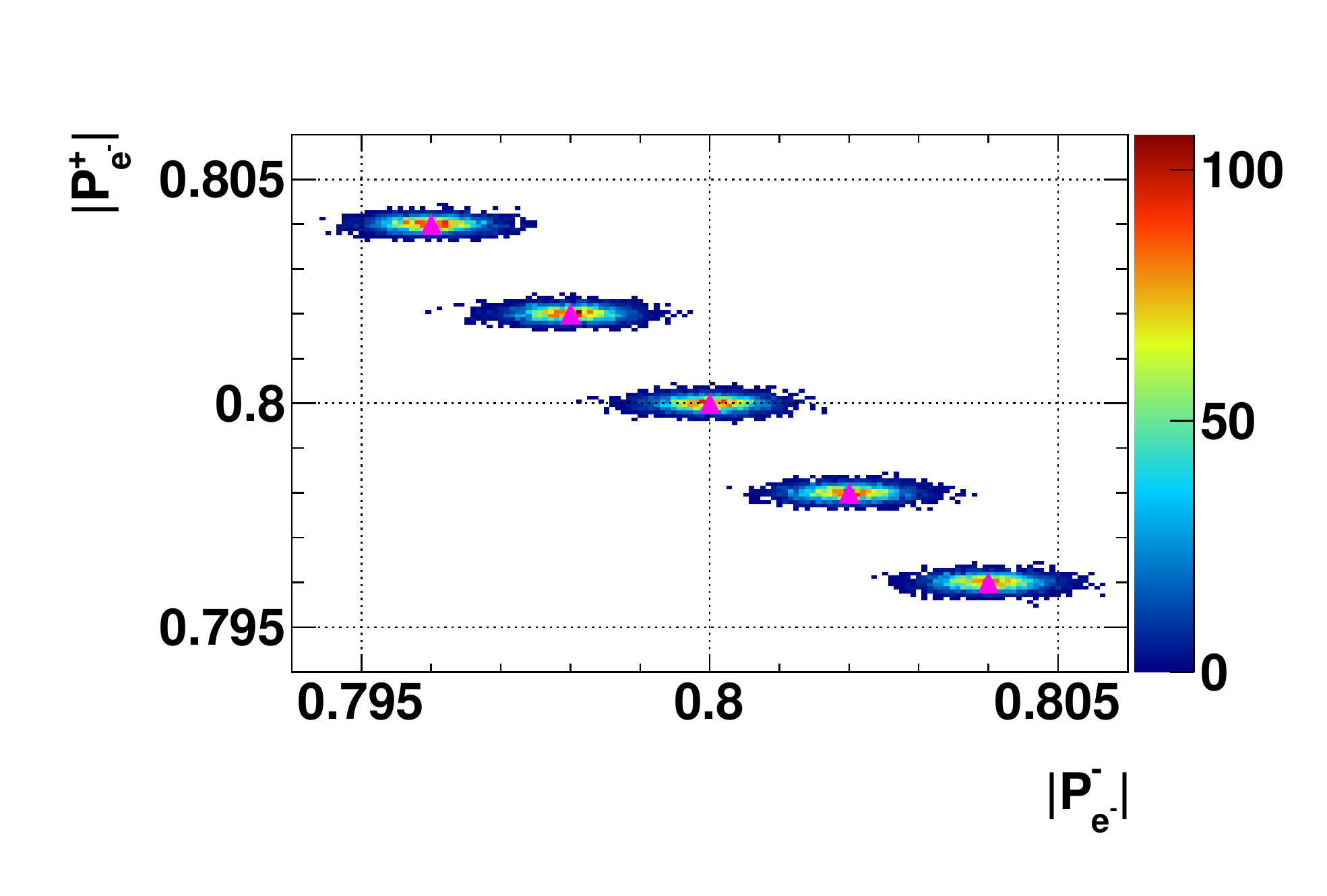}
\includegraphics[width=0.49\textwidth]{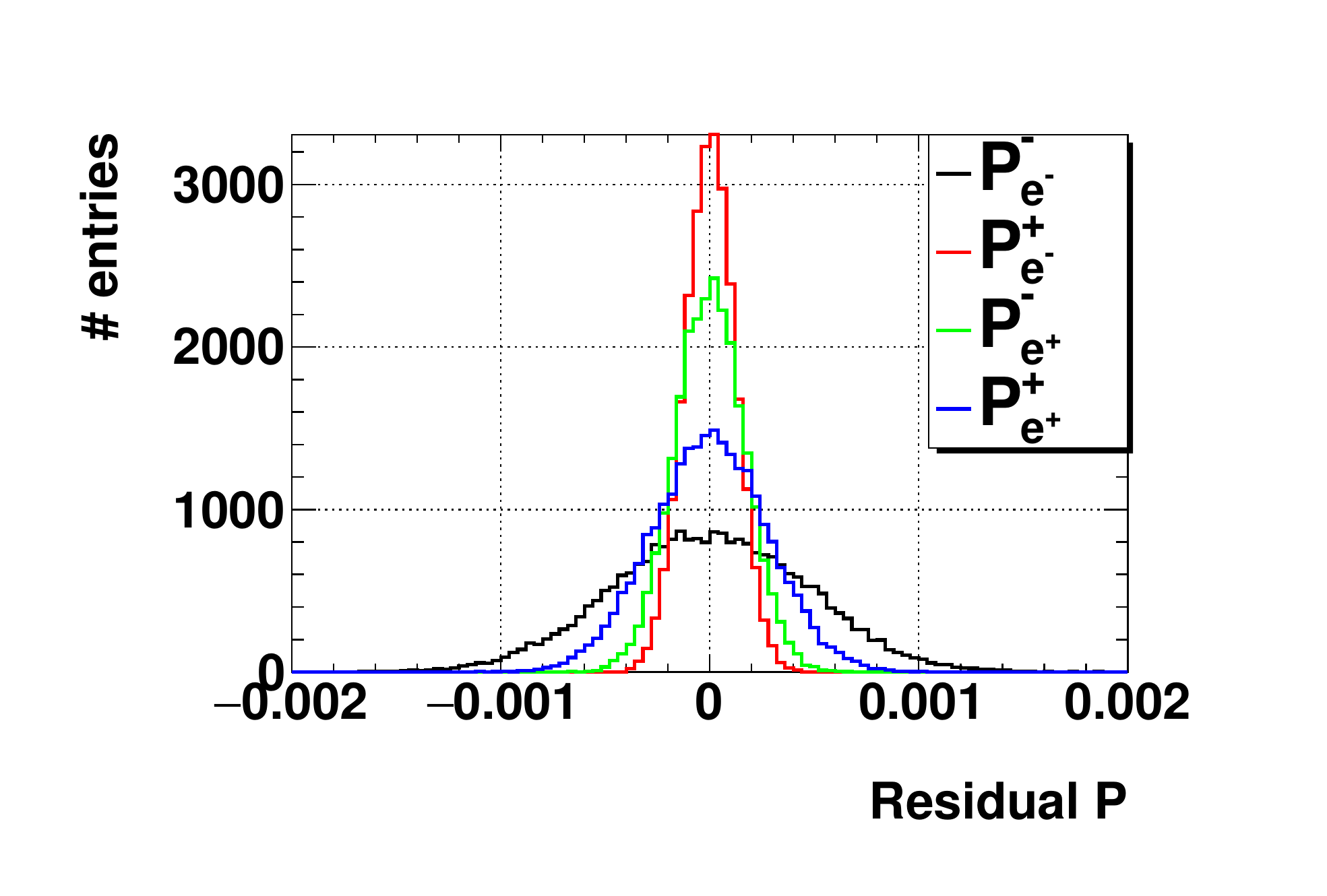}
\caption{Results for 5 different absolute polarization deviations after helicity reversal. \textit{Left}: the correlation between the two helicity configuration of the electron beam from the toy measurements. The magenta triangles corresponds to actual polarization values. \textit{Right}: The residual of the measured polarization from the actual polarization combined for all 5 polarization discrepancy.}
\label{fig:nonperfecthelicityreversal}
\end{figure}

As seen in the left plot of Fig.~\ref{fig:nonperfecthelicityreversal}, the results of $\chi^{2}$-fit fluctuating within their uncertainties around the actual beam polarization values, displayed as magenta triangles. As seen in the right plot of Fig.~\ref{fig:nonperfecthelicityreversal}, all deviations from the actual polarization only occur due to the statistical uncertainty. A change in the uncertainty in comparison to equal absolute polarization is not noticeable. Therefore, the assumption that a non-perfect helicity reversal has no influence on the polarization determination is justified because the absolute polarizations are correctly determined and there is also no change in the uncertainty. Thus, the new unified approach compensates for a non-perfect helicity reversal.\\

\section{Improvement by Constraints from Polarimeter Measurement}
\label{sec:polarimeterconstraint}

In a simplified approach, the effects of using the polarimeter measurement as an additional constraint were studied. This is just a first step to demonstrate the proof of principle. In this approach, spin transport was neglected and it was pretended that the polarimeter could measure the polarization direct at the IP with the relative nominal precision of $\Delta P/P = 0.25\%$. Furthermore, a Gaussian distribution of the polarimeter measurement is assumed with the nominal absolute polarization values $\left|P_{e^{-}}\right| = 80\%,\left|P_{e^{+}}\right| = 30\%$ as mean value and the nominal uncertainty $\Delta P$ as deviation.\\
For the implementation in the $\chi^{2}$ minimization, the squared pull terms of the 4 polarization parameters and the polarimeter measurement were additionally added to the existing $\chi^{2}$ function, as seen in Fig.~\ref{eq:chiwithconstraint}

\begin{align}
	\chi^{2} &=
	\underbrace{\sum_{\text{process}}{
	\left(\vec{\sigma}_{\text{data}} - \vec{\sigma}_{\text{theory}}\right)^{T}\Xi^{-1}
	\left(\vec{\sigma}_{\text{data}} - \vec{\sigma}_{\text{theory}}\right)
	}}_{\text{Contribution from the cross section measurement}} + \underbrace{\sum_{P}{\left[\frac{\left(P_{e^{\pm}}^{\pm} - \mathcal{P}_{e^{\pm}}^{\pm}\right)^{2}}{\Delta\mathcal{P}^{2}}\right]}}_{\text{Polarimeter term}}
\label{eq:chiwithconstraint}	
\end{align}

\begin{figure}[htbp]
\centering
\begin{minipage}{0.52\textwidth}
\includegraphics[width=\textwidth]{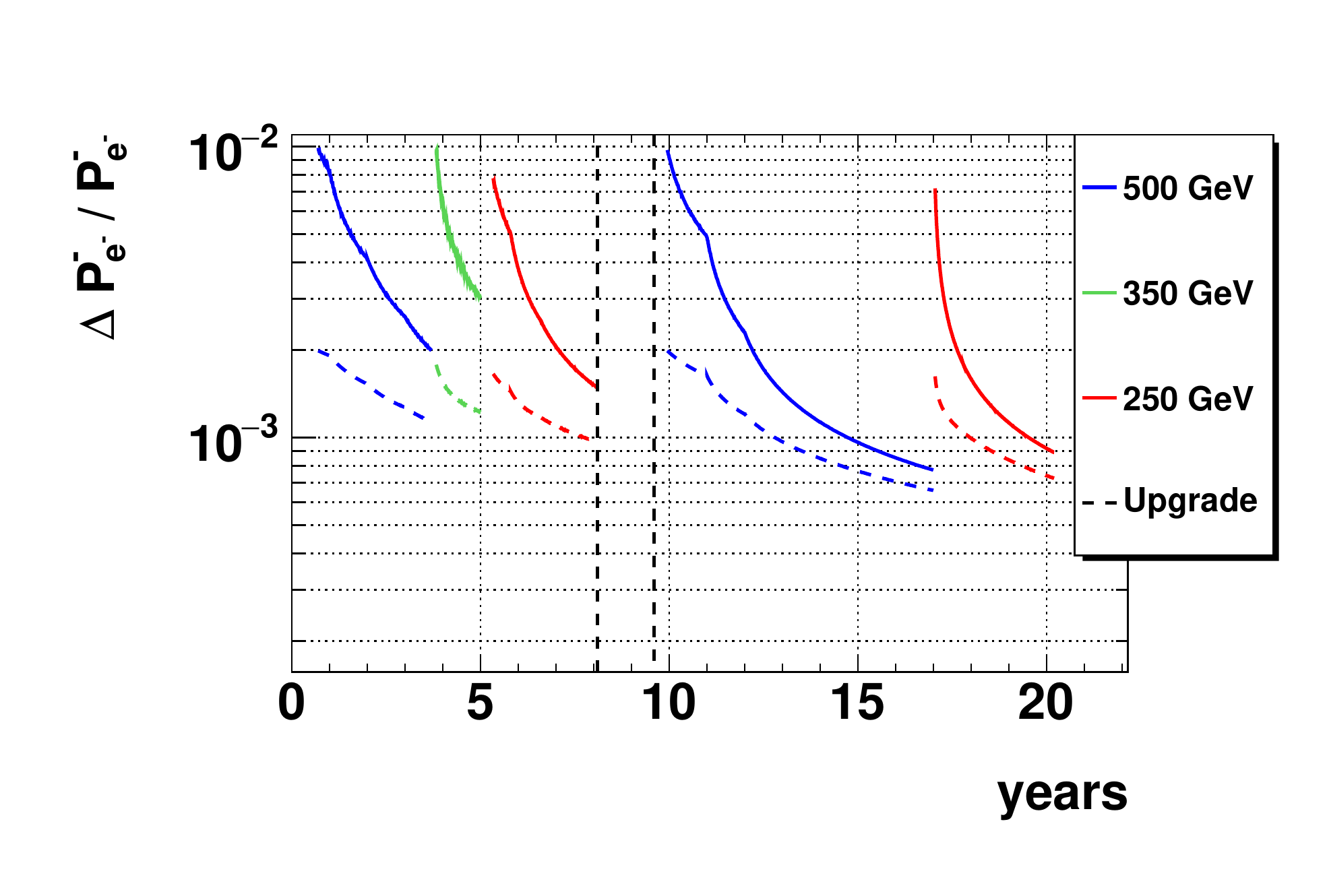}
\end{minipage}\hfill
\begin{minipage}{0.4\textwidth}
\begin{tabular}{|c|c|c|c||c|c|}\hline
$E$ & $\textbf{500}$ & $\textbf{350}$ & $\textbf{250}$ & $\textbf{500}$ & $\textbf{250}$\\\hline
$\mathcal{L}$ & $\textbf{500}$ & $\textbf{200}$ & $\textbf{500}$ & $\textbf{3500}$ & $\textbf{1500}$ \\\hline
\multicolumn{6}{|c|}{\textbf{Without Constraint}}\\\hline
$P_{e^{-}}^{-}$ & \textcolor{red}{$0.2$} & \textcolor{red}{$0.3$} & $0.1$ & $0.08$ & $0.09$\\\hline
\multicolumn{6}{|c|}{\textbf{With Constraint}}\\\hline
$P_{e^{-}}^{-}$ & \textcolor{blue}{$0.1$} & \textcolor{blue}{$0.1$} & $0.1$ & $0.07$ & $0.07$\\\hline
\end{tabular}
\end{minipage}
\caption{The improvement on the polarization precision for the left-handed electron beam by using the polarimeter constraint displayed for the H-20\cite{ILCrunningscenario} scenario.}
\label{figtab:polarimeterconstraint}
\end{figure}

Here, $P_{e^{\pm}}^{\pm}$ are the 4 polarization parameters used also in the other part of the $\chi^{2}$ function and $\mathcal{P}_{e^{\pm}}^{\pm}$ are the corresponding polarimeter measurement. Here the advantage of the parametrization becomes clear because the free polarization parameter can be directly compared with the polarimeter measurement without any further conversion of the parameters. $\Delta\mathcal{P}$ is the uncertainty on the polarimeter measurement.\\
The results for the H-20\cite{ILCrunningscenario} scenario can be seen in Fig.~\ref{figtab:polarimeterconstraint}. With the constraint, the overall polarization precision can be improved for all runs. This is of particular importance for runs with less integrated luminosity. As described earlier, the precision goal of a permille-level can not be reached in such runs, which is now possible due to the polarimeter constraint.\\

As a next step, the constraint has to be modified to a more realistic scenario. First of all, the polarimeters do not measure the polarization at the IP, thus the measurement has to be extrapolated. However, the information of the polarimeters can also be included to compensate for misalignments in the BDS (Spin tracking) and for beam collision effects.\cite{spintracking}\\

\section{Conclusion and Outlook}

Polarization provides a deep insight in the chiral structure of the standard model and beyond. Therefore, a permille-level precision of the luminosity-weighted average polarization at the IP is required. With the new unified approach, all suitable cross section measurements as well as the constraints of the polarimeter measurement are combined by using a overall $\chi^{2}$ minimization\\
With this approach a statistical precision of a permille-level is achievable. Furthermore, the impact of time-dependent systematic uncertainties can be reduced due to a fast helicity reversal, while a non-perfect helicity reversal has no impact on the statistical precision. Thereby, the further improvement due to polarimeter constraints is particular important for the low integrated luminosity.\\
A further topic will attend to the time-dependence of the beam polarization. Currently there is always the assumption of a time-independent beam polarization. But with the H-20\cite{ILCrunningscenario} scenario in mind, it is not realistic to assume that the polarization stays constant over a time of up to 7 years. Thus,  potential time-dependencies also have to be corrected in the polarimeter constraint. An important point is that it is straight-forward to include such effects in the new unified approach.\\
In order to gain a realistic and precise description of the polarization uncertainty, it is important to consider an accurate estimation of the systematic quantities (selection efficiency $\varepsilon$ and background estimation $B$), their uncertainty ($\Delta B$, $\Delta\varepsilon$, $\Delta\mathcal{L}$) and correlations. Although systematic quantities are fully implemented in the $\chi^{2}$ algorithm, their actual value still has to be determined.\\
Using the angular information of a process, the precision on the polarization can be further improved. This is achieved by implementing differential cross sections within the $\chi^2$-method.\\

\end{document}